\RequirePackage{amsmath}
\documentclass[runningheads]{llncs}

\usepackage{graphicx}
\usepackage{todonotes}
\usepackage{bbding}
\usepackage{booktabs}
\usepackage{multirow,bigdelim}
\usepackage{color}
\usepackage{xr}
\usepackage[colorlinks = true,
            linkcolor = blue,
            urlcolor  = blue,
            citecolor = blue,]{hyperref}
\usepackage{textcomp}
\usepackage{gensymb}
\usepackage{amssymb}

\begin{document}

\title{Multiple Sclerosis Lesion Segmentation - \texorpdfstring{\\ A Survey of Supervised CNN-Based Methods}{A Survey of Supervised CNN-Based Methods}}

\titlerunning{MS Lesion Segmentation: Supervised CNN-Based Methods}

\author{Huahong Zhang\inst{1} \and
Ipek Oguz\inst{1 (}\Envelope\inst{)}
}


\authorrunning{H. Zhang et al.}

\institute{Vanderbilt University, Nashville, TN 37235, USA \\
\email{\{huahong.zhang,ipek.oguz\}@vanderbilt.edu} }

\maketitle

\begin{abstract}
Lesion segmentation is a core task for quantitative analysis of MRI scans of Multiple Sclerosis patients. The recent success of deep learning techniques in a variety of medical image analysis applications has renewed community interest in this challenging problem and led to a burst of activity for new algorithm development.
In this survey, we investigate the supervised CNN-based methods for MS lesion segmentation. We decouple these reviewed works into their algorithmic components and discuss each separately. For methods that provide evaluations on public benchmark datasets, we report comparisons between their results.

\end{abstract}

\section{Introduction}
Multiple Sclerosis (MS) is a demyelinating disease of the central nervous system. For monitoring the disease course, focal lesion quantification using magnetic resonance imaging (MRI) is commonly used. Recently, deep learning has achieved great success in the computer vision community \cite{krizhevsky_imagenet_2012} as well as in medical image analysis tasks, and has been applied to MS lesion segmentation \cite{carass_longitudinal_2017,valverde_improving_2017,zhang_multiple_2019}. Accurate lesion segmentation is valuable for clinical application and subsequent analysis (e.g., \cite{zhang2020robust}).

In this survey, we focus on the methods which are CNN-based, and we review about 100 papers that were published until late October 2020. However, we do not intend to review these papers exhaustively.  
Instead, we break down the reviewed segmentation pipelines into algorithmic components such as data augmentation and network architecture and compare the representative advances for each component in Sec.~\ref{sec:review_of_methods}. 
This is different from the previous surveys of MS segmentation methods (e.g., \cite{kaur_state---art_2020,danelakis_survey_2018}).

Unsupervised learning methods are not included in this survey since an excellent comprehensive survey of these is already provided by Baur \emph{et al.}\cite{baur_autoencoders_2020}. As a general note, supervised methods tend to perform better than unsupervised methods for MS lesion segmentation, given the difficulty of the task (\cite{carass_longitudinal_2017,commowick_objective_2018}). 

To identify the articles to include in this review, we conducted a Google Scholar search using the keywords ``Multiple Sclerosis + lesion + segmentation + neural network''. To ensure we include important advances, we also went through the references cited in each reviewed paper. In addition, papers who cite  publicly available MS datasets (e.g., \cite{carass_longitudinal_2017,commowick_objective_2018}) were considered. 

\section{Review of Methods} \label{sec:review_of_methods}

\subsection{Data Pre-processing}

Pre-processing is a common practice in the reviewed papers. It usually includes skull stripping (e.g., BET\cite{smith_fast_2002}), bias field correction (e.g., N4ITK\cite{tustison_n4itk:_2010}), rigid registration and intensity normalization. 
Rigid registration is used to register between different MRI modalities acquired during a scanning session, between scans of a given patient acquired at different time points, as well as between different subjects of a study or to standard template spaces (e.g., MNI152 \cite{fonov2009unbiased}).

For intensity normalization, there are two popular approaches: 1) histogram matching (e.g., \cite{vaidya2015longitudinal,birenbaum_longitudinal_2016}) and 2) normalizing data to a specific range. For this latter approach, it is common to enforce zero mean and unit variance (whitening-like, e.g., \cite{havaei_hemis_2016,valverde_improving_2017,valverde_one-shot_2018}) or fit the intensities into the range [0, 1] (e.g., \cite{feng_self-adaptive_2018,ulloa_circular_2019,ghafoorian_transfer_2017}). Some pipelines (e.g., \cite{vaidya2015longitudinal,birenbaum_longitudinal_2016}) also chose to only preserve the values between a range (e.g., 1\textsuperscript{st} and 99\textsuperscript{th} percentile) before normalization to minimize the effect of intensity outliers. Further combinations of these two approaches are also possible. Ravnik \emph{et al.} \cite{ravnik_dataset_2018} argued that in a same-scanner, homogeneous situation, using (whitening-like) intensity normalization, histogram standardization, or both all achieved similar results, and indeed these had no statistically significant improvement over even no normalization at all. But in a multi-scanner, heterogeneous dataset, only using normalization is slightly better than using both, and all performed statistically better than no pre-processing. Other advanced pre-processing techniques (e.g., white stripe \cite{shinohara2014statistical}) can also be considered.


\subsection{Data Representation} \label{data_representation}

For feeding the deep neural network, the input data are usually represented as patches of raw MRI images since whole images are too large.
These patches can be 2D, 3D, or any format in between.
Choosing the format of patches is an important design decision and will affect the performance of networks. Whether the data is of isotropic resolution needs to be considered for making this decision \cite{isensee_nnu-net_2019}.

With \textbf{2D} (slice-based) patches (e.g., \cite{roy_multiple_2018,aslani_multi-branch_2018,zhang_ms-gan:_2018}), the advantage is that there are far fewer network parameters to train, and therefore, these tend to be not as prone to over-fitting compared to 3D networks. However, contextual information along the third axis is missing. 
In contrast, \textbf{3D} approaches (e.g., \cite{brosch_deep_2016,valverde_improving_2017,hashemi_asymmetric_2019,feng_self-adaptive_2018}) take advantage of the local contextual information, but they do not have any global context. Compared to 2D methods, due to the small size used in 3D patches, the long-distance spatial information cannot be preserved. Further, for processing 3D data, the networks are computationally expensive, need more parameters, and are prone to over-fitting. 


\paragraph{\bf Multi-view/2.5D.}
To obtain a balance between 2D and 3D representations, Birenbaum \emph{et al.} \cite{birenbaum_longitudinal_2016} proposed to use multi-view data, in which three orthogonal 2D views passing through the same voxel are input to the network jointly. 
On the other hand, Aslani \emph{et al.} \cite{aslani_multi-branch_2018} and Zhang \emph{et al.}~\cite{zhang_multiple_2019} used 2.5D by extracting slices from different planes, which are used to train the networks independently from each other. Zhang \emph{et al.} \cite{zhang_multiple_2019} also use ``stacked slices'' by stacking 2D patches along the third axis, which generates thinner but larger 3D input than normal 3D patches. Using 2.5D and prediction fusion, the networks are able to learn and utilize global information along all three axes.

\subsection{Data preparation} \label{data_preparation}
\paragraph{\bf Candidate Extraction.} 

To extract input patches from the raw data, an intuitive way is to move a sliding window voxel-by-voxel throughout the raw MRI volume. However, this approach will generate a lot of similar patches and leads to class imbalance.
The strategy used to alleviate these problems is slightly different between fully convolutional network (FCN) and non-FCN methods (as discussed in Sec \ref{network}). This distinction is mainly due to the labels of patches for non-FCN methods being single scalars, which are more vulnerable to class imbalance. Furthermore, FCN methods usually solve class imbalance with loss functions.

For \textbf{non-FCN} methods, 
Birenbaum \emph{et al.} \cite{birenbaum_longitudinal_2016} proposed to apply a probabilistic WM template and choose the patches whose center is a voxel with high intensity in FLAIR and high probability in WM template. The method does not need lesion maps, so it can be used in both training and test phases to reduce the computation burden.
Valverde \emph{et al.} \cite{valverde_improving_2017,valverde_one-shot_2018} randomly under-sample the negative class to make a balanced dataset. 
Kazancli \emph{et al.} \cite{kazancli_multiple_2018} considered strategies of random sampling and sampling around the lesions, with the latter providing better model performance.
Ulloa \emph{et al.} \cite{ulloa_circular_2019} augmented data from the lesion class to balance the dataset instead of down-sampling the non-lesion class. Also, they use circular non-uniform sampling, which allows greater contextual information with a radial extension. They further presented a stratified sampling method \cite{ulloa_improving_2020}. Among voxels not labeled as lesions, a portion $p$ of the candidates is extracted from the neighborhood of lesions and the remaining $1-p$ from the remaining voxels.

For \textbf{FCN} methods, Kamnitsas \emph{et al.} \cite{kamnitsas_efficient_2017} extract the patches with a 50\% probability of being centered on a foreground or background voxel. Feng \emph{et al.} \cite{feng_self-adaptive_2018} centered the patches at a lesion voxel with a probability $p=0.99$.
Many methods (e.g., \cite{roy_multiple_2018,aslani_multi-branch_2018,zhang_multiple_2019,aslani_deep_2019}) only extract patches with at least one lesion voxel.




\paragraph{\bf Augmentation.} 
After extracting the patches, data augmentation techniques can be applied. Commonly used augmentations include random flip, random rotation, and random scale. Usually, the rotation is 2D and the angle is $n\times90\degree (n\in \mathbb{Z})$. However, a random angle (e.g., \cite{birenbaum_longitudinal_2016,aslani_deep_2019}) or a 3D rotation have also been used. Sometimes random noise and random ``bias field'' can be added \cite{kruger2020fully}.
Additionally, Salem \emph{et al.} \cite{salem_multiple_2019} suggested synthesizing lesions may be helpful as an augmentation. 


\paragraph{\bf Label Processing/Denoising.}

For training and evaluating models, the label can be a scalar representing whether the central voxel of the input patch is a lesion or not, or it can be the lesion map the same size as the input, depending on the network type (discussed in Sec.~\ref{network}). Either the scalar or the lesion map is extracted from the expert delineations. These unavoidably contain noise, which usually comes from the lesion borders \cite{kats_soft_2019}. 
Also, some datasets (e.g., \cite{carass_longitudinal_2017,commowick_objective_2018}) are delineated by more than one expert, and inter-rater variability needs to be addressed. While a simple majority vote can be used, STAPLE \cite{warfield2004simultaneous} and its variation \cite{akhondi2014logarithmic} are very common. 
On the other hand, a few methods (e.g., \cite{zhang_multiple_2019}) treat delineations from different experts as different samples. In other words, they train networks with the same input patch but different labels. 

Even consensus delineations still contain noise that can be further mitigated. Roy \emph{et al.} \cite{roy_multiple_2018} generated training memberships (labels) by convolving the binary segmentations with a $3\times 3$ Gaussian kernel to get a softer version of boundaries.
Kats \emph{et al.} \cite{kats_soft_2019} proposed to assign fixed soft labels to the voxels near the lesions by 3D morphological dilation. Even though they define the soft-Dice loss for this purpose, an easier solution may be to generate the soft version of labels at the data preparation stage. Cohen \emph{et al.} \cite{cohen_learning_2020} further proposed to learn soft labels instead of fixed values, and then apply the soft-STAPLE algorithm \cite{kats2019soft} to fuse the masks. 

\paragraph{\bf Extra Information.}

Extra information, for example, spatial locations or intermediate results provided by other models, can also be used as input to CNNs.
Ghafoorian \emph{et al.} \cite{ghafoorian_location_2017} incorporated eight spatial features, with dense features extracted by convolutional layers and fed into fully connected layers. La Rosa \emph{et al.} \cite{la_rosa_shallow_2018} provided the CNNs with the probability maps generated by a Bayesian partial volume estimation algorithm \cite{fartaria2017segmentation}.

\subsection{Network Architecture} \label{network}

The network architecture plays a very important role in deep learning. Many works focus on crafting the structure to improve the segmentation performance.

\paragraph{\bf Network Backbones.}


After Krizhevsky \emph{et al.} \cite{krizhevsky_imagenet_2012} won the ImageNet 2012 challenge, Convolutional Neural Networks (\textbf{CNN}s) became very popular and have been successfully applied to medical imaging problems \cite{carass_longitudinal_2017,commowick_objective_2018}.
For MS lesion segmentation with CNN, the early methods use voxel-wise classification to form the lesion segmentation map \cite{birenbaum_longitudinal_2016,kazancli_multiple_2018,valverde_improving_2017}. 
A typical network of this type consists of a few convolutional layers followed by 2-3 fully connected layers (also called Multilayer Perceptron, MLP) for the final prediction.
The input is an image patch, and the networks are trained to classify whether the central voxel of this patch corresponds to a lesion. While these methods have outperformed conventional methods, they have disadvantages: 1) lack of spatial information as only small patches are used; 2) computational inefficiency due to the repetition of similar patches.

Kang \emph{et al.} \cite{kang_fully_2014} introduced the fully convolutional neural network (\textbf{FCN}/FCNN) for the segmentation task. FCNs do not need to make the lesion prediction by classifying the central voxel of each patch. Instead, they directly generate lesion maps of the same (or similar) size as the input images. However, due to the successive use of convolutional and pooling layers, this approach produces segmentations at a lower resolution. Long \emph{et al.} \cite{long_fully_2015} preserve the localization information from low-level features and contextual information from high-level features by adding skip connections. Applications of FCNs for MS lesion segmentation include \cite{mckinley2016nabla,brosch_deep_2016,roy_multiple_2018} and some of these methods take advantage of shortcut connections \cite{mckinley2016nabla,brosch_deep_2016}. 

Ronneberger \emph{et al.} \cite{ronneberger_u-net:_2015} then used a symmetrical u-shape network called \textbf{U-Net} to combine features. The network has an encoder-decoder structure and adds shortcut connections between corresponding layers of the two parts. 
The pooling operations in the encoder are replaced by upsampling operations in the decoder path. Many recent MS segmentation methods \cite{salehi_tversky_2017,aslani_multi-branch_2018,feng_self-adaptive_2018} are based on the original U-Net or slight modifications thereof.



\paragraph{\bf Variations of U-Net.} 


For CNNs with voxel-wise prediction or FCN methods that are not U-Net-based, the network structures are quite flexible. As for pipelines that use the U-Net, the most common modification is to introduce some crafted modules (residual block, dense block, attention module, etc.). These new modules can be added to replace the convolutions or between the shortcut connections.
Aslani \emph{et al.} \cite{aslani_multi-branch_2018,aslani_deep_2019} presented a U-Net-like structure with convolutions replaced by residual blocks in the encoder path. 
Hashemi \emph{et al.} \cite{hashemi_asymmetric_2019} and Zhang \emph{et al.} \cite{zhang_multiple_2019} adopted the Tiramisu network, which replaces the convolution layers with densely connected blocks (skip connection between layers).
Hu \emph{et al.} \cite{hu_acu-net_2020} presented a context-guided module to expand the perception field and utilize contextual information. 
Vang \emph{et al.} \cite{vang_synergynet_2020} augmented the U-Net with the Mask R-CNN framework.

\underline{Attention Module.} Attention mechanism has been researched in many works for MS lesion segmentation. In general, it can be divided into spatial attention, channel attention, and longitudinal attention.
Zhang \emph{et al.} \cite{zhang_rsanet_2019} presented a recurrent slice-wise attention network for 3D CNNs. By performing slice-wise attention from three orientations, they reduced the memory demands for 3D spatial attention.
Hu \emph{et al.} \cite{hu_acu-net_2020} included 3D spatial attention blocks in the decoding stage.
Durso-Finley \emph{et al.} \cite{durso-finley_saliency_2020} used a saliency-based attention module before the U-Net structure to make the network realize the difference between pre- and post-contrast T1-w images and thus focus on the contrast-enhancing lesions. 
Hou \emph{et al.} \cite{hou_cross_2019} proposed a cross-attention block that combines spatial attention and channel attention.
Zhang \emph{et al.} \cite{zhang2020efficient} extend their folded (slice-wise) attention to the spatial-channel attention module. They use four permutations (corresponding to four dimensions) to build four small sub-afﬁnity matrices to approximate the original affinity matrix. In such a case, the original affinity matrix is regularized as a rank-one matrix and the computational burden is alleviated.
Gessert \emph{et al.} \cite{gessert_multiple_2020} introduced attention-guided interactions to enable effective information exchange between the processing paths of the two time points.



\paragraph{\bf Multi-task Networks.}

Narayana \emph{et al.} \cite{narayana_multimodal_2018} performed segmentation of brain tissue and T2 lesions at the same time. McKinley \emph{et al.} \cite{mckinley_simultaneous_2019} illustrated that the inclusion of additional tissue classes during the segmentation of lesions is helpful for MS segmentation. Duong \emph{et al.} \cite{duong_convolutional_2019} trained the networks with data from many different tasks, making the trained CNN usable for multiple tasks without tuning and thus more applicable in a clinical context.

\subsection{Multiple Modalities, Timepoints, Views and Scales}\label{sec:multi-sources}


In the context of MS lesion segmentation, the incorporation of multi-modality, multi-timepoints, multi-view and multi-scale data are similar. The data from different sources have to be fused at some point of the pipeline: input, feature map, and/or output. Fusing at the input can be simple concatenation along channels while fusing the output is roughly equivalent to making an ensemble to reach consensus. Fusing the features usually needs parallel paths and interaction between paths, which typically happens in the encoder path in a U-Net-like structure.

\paragraph{\bf Multi-modalities.} The commonly used MRI sequences for MS white matter lesion segmentation include T1-weighted (T1-w), T2-weighted (T2w), proton density-weighted (PD-w) and fluid attenuated inversion recovery T2 (FLAIR). 

Narayana \emph{et al.} \cite{narayana_are_2020} evaluated the performance of U-Net when it is trained with different combinations of modalities on a large multi-center dataset. They concluded that using all the modalities, especially with FLAIR, achieved the best performance. A similar conclusion can be found in \cite{brosch_deep_2016} and other works that use multiple MRI sequences as input. For fusing the different modalities, 
Roy \emph{et al.} \cite{roy_multiple_2018} use parallel pathways for processing different modalities and then concatenate the features along the channels (only once). Aslani \emph{et al.} \cite{aslani_multi-branch_2018} use parallel encoder paths to process different modalities, and they fuse the different modalities after each convolutional block. 
Zhang \emph{et al.} \cite{zhang_ms-gan:_2018} also use a similar strategy. 
Zhang \emph{et al.} \cite{zhang_multiple_2019} fuse the patches from different sequences before feeding into the network.


Multi-modality methods are usually trained on a specific set of modalities and thus require these sequences to be available at the test phase, which can be limiting. 
To deal with missing modalities, Havaei \emph{et al.} \cite{havaei_hemis_2016} propose to use parallel CNN paths to learn the embeddings of different input sequences into a single latent space for which arithmetic operations are well deﬁned. 
They randomly drop modalities during training.
As such, any subset of available modalities can be used as input at test time.
Feng \emph{et al.} \cite{feng_self-adaptive_2018} also use random ``dropout'' of modalities but substitute the missing modalities with background values.


\paragraph{\bf Multi-timepoints.} Longitudinal studies are common in MS, but the ongoing inflammatory disease activity complicates the analysis of longitudinal MRIs as new lesions can appear and existing lesions can heal between scans. To improve individual \textbf{segmentation performance}, Birenbaum \emph{et al.} \cite{birenbaum_longitudinal_2016} propose to process the two time-points individually with a Siamese architecture, where the parallel paths share weights, and then concatenate the features for classification. 
Denner \emph{et al.} \cite{denner_spatio-temporal_2020} argue that this late-fusion strategy \cite{birenbaum_longitudinal_2016} does not properly take advantage of learning from structural changes and they propose two complementary networks for multi-timepoints. The longitudinal network fuses the two time-points early to implicitly use the structural differences. The multi-task network is trained to learn the segmentation with an additional task of deformable registration between the two time-points, which explicitly guides the network to use spatio-temporal information.

To identify \textbf{lesion activity}, Placidi \emph{et al.} \cite{placidi_automatic_2019} simply segment the lesions at the two time-points independently and register the previous examination to the current examination to compare the segmentations. However, comparing differences between time-points relies on high similarity between scans and requires highly accurate registration. 
McKinley \emph{et al.} \cite{mckinley_automatic_2020} introduce a segmentation confidence for comparing the lesion segmentation between timepoints. Comparisons are based on the ``confident'' lesions of each timepoint.
Kruger \emph{et al.} \cite{kruger2020fully} fed two timepoints into the same encoder (share weights) and the feature maps are concatenated after each residual block before going to the corresponding decoder block. 
Salem \emph{et al.} \cite{salem_fully_2020} use cascaded networks for detecting new lesions. The first network learns the deformation field between the baseline and follow-up images. The second network takes the two images and the deformation field, and outputs the segmentation. 
To assist the network in learning to detect new and enlarging T2w lesions, Sepahvand \emph{et al.} \cite{sepahvand2020cnn} illustrate an attention-like mechanism. They multiply the multi-modal MRI at the reference (in contrast to follow-up) with subtraction images, which acts as the attention gate. Then the product is concatenated with the lesion map at the reference to feed the network.
Gessert \emph{et al.} \cite{gessert_4d_2020} propose convolutional gated recurrent units for temporal aggregation. The units are inserted into the bottleneck and skip connections of U-Net. In another work \cite{gessert_multiple_2020}, the same team process two timepoints with parallel encoder paths that interact with each other using attention modules. In such a scenario, the attention mechanism functions similarly to masking early time-points. 


\paragraph{\bf Multi-views.} As discussed in Sec. \ref{data_representation}, multi-view data can be utilized as data representation between 2D and 3D. To handle multi-view data, Birenbaum \emph{et al.} \cite{birenbaum_longitudinal_2016} processed different views by parallel sub-networks and then concatenated the features to feed the fully connected layers.
McKinley \emph{et al.} \cite{mckinley2016nabla} integrate three networks for the three views and the outputs are averaged.
Zhang \emph{et al.} \cite{zhang_multiple_2019} use one network for different views, i.e., the network parameters are shared between views.
Shachor \emph{et al.} \cite{shachor_mixture_2020} propose a gated model Mixture of Views (MoV) to fuse different views. 


\paragraph{\bf Multi-scales.} The networks based on FCN and U-Net inherently incorporate multiple scales. However, explicit use of multi-scales may also be useful. For non-FCN networks, Kamnitsas \emph{et al.} \cite{kamnitsas_efficient_2017} propose to use two parallel paths, one for full resolution and another for a lower resolution; these two paths are fused before fully connected layers.
As for U-Net-based methods, Wang \emph{et al.} \cite{wang_ensemble_2018} argue that different types of segmentation biases may be generated by networks of different input sizes. To address this issue, they train 3 networks with different input sizes and use another stage for fusing the results. Hou \emph{et al.} \cite{hou_cross_2019} feed multi-scale input (original and downsampled patches) to the first three layers and they aggregate the multi-scale outputs from the last three layers to make the final prediction.
Hu \emph{et al.} \cite{hu_acu-net_2020} use one input and average the multi-scale outputs.

\paragraph{\bf Others.} For training data with annotations from multiple experts, Vaidya \emph{et al.} \cite{vaidya2015longitudinal} train two separate networks using the same images but different delineations from two experts. The outputs of the two networks are averaged to get the final prediction. Zhang \emph{et al.} \cite{zhang2020learning} present a segmentation network to estimate the ground truth segmentation and an annotator network to estimate the characteristics of each expert, which can be viewed as a translation of STAPLE to CNN. 

No-new-UNet (nnU-Net) \cite{isensee_nnu-net_2019} is a multi-architecture framework that adaptively chooses an ensemble from 2D, 3D, 3D cascaded networks. 

\subsection{Loss Functions and Regularization}

For MS lesions segmentation, which is usually binary, the most commonly used loss function is the Binary Cross-Entropy (\textbf{BCE}, e.g.,  \cite{birenbaum_longitudinal_2016,aslani_deep_2019}).
Other losses such as \textbf{L2 loss} (e.g., \cite{brosch_deep_2016,zhang_multiple_2019}) are also explored. As class imbalance exists within the dataset, the original losses can be weighted based on the probability (prevalence) of each class. The class with the lower probability (i.e., lesion) is compensated with a higher weight. Brosch \emph{et al.} \cite{brosch_deep_2016} implicitly weight lesion voxels and non-lesion voxels by calculating the \textbf{weighted L2 loss} of the lesion voxels and non-lesion voxels. Feng \emph{et al.} \cite{feng_self-adaptive_2018} used \textbf{weighted BCE} with a lesion/non-lesion ratio of 3 to 1. \textbf{Focal loss} \cite{lin_focal_2017}, as the generalization of BCE, was proposed not only to weight the lesion class but also to give more importance to hard examples (e.g., \cite{zhang_multiple_2019,hashemi_asymmetric_2019}).


For FCN-based methods, since the labels/outputs are patches, region-based losses are used to address the intra-patch class imbalance.
Milletari \emph{et al.} \cite{milletari_v-net:_2016} proposed the \textbf{Dice ($F_1$) loss} to balance between precision and recall equally. 
\textbf{Tversky loss}  \cite{salehi_tversky_2017,hashemi_asymmetric_2019} is the generalization of the Dice loss and $F_\beta$ loss (such that $\beta=1$ is the Dice loss). 
The networks trained with higher $\beta$ have higher recall and lower precision. Based on this property, Ma \emph{et al.} \cite{ma2020ensembling} trained individual models with high $\beta$ to make up an ensemble. Their assumption is that diverse low-precision-high-recall models tend to make different false-positive errors but similar consistent true-positives. Thus the false-positive errors can be canceled out by aggregating the predictions of the ensemble.
Further, the \textbf{Focal Tversky loss} (e.g., \cite{hu_acu-net_2020}) is the generalization of the Tversky loss. It is similar to Focal loss, which is capable of focusing on mislabeled samples and minority samples by controlling the parameters.


Combining loss functions and introducing domain-specific regularization are helpful in some cases. 
McKinley \emph{et al.} \cite{mckinley2016nabla} calculated the 25\textsuperscript{th} percentile of the intensity within the lesion mask and weighted the loss function from these voxels higher than others.  
Zhang \emph{et al.} \cite{zhang_ms-gan:_2018} proposed to use Generative Adversarial Network (GAN) architecture to provide an additional discriminator-based constraint. They use a combination of BCE loss, Dice loss, L1 loss and GAN-loss. 
Additionally, loss functions have been proposed to help the networks with uncertainty analysis \cite{mckinley_simultaneous_2019}, domain adaptation \cite{baur_semi-supervised_2017,ackaouy_unsupervised_2020,aslani_deep_2019} and other goals. 

\subsection{Implementation}

To train the networks, a simple strategy is using fixed epochs. However, early stopping is a common practice to avoid over-fitting. Thus, the training dataset is divided into a fixed training subset and validation subset, or k-fold cross-validation can be utilized \cite{birenbaum_longitudinal_2016,hashemi_asymmetric_2019,zhang_multiple_2019}. The ``best'' models are chosen based on model performance (e.g., Dice score) on the validation set. To provide a fair evaluation, the test set is usually held-out from the training/validation data and different scans of the same patient should not be placed into different datasets.

For optimizing the parameters, stochastic gradient descent (SGD) and its variations are used. To avoid oscillation in local optima, the momentum variable was introduced (e.g., \cite{hu_acu-net_2020}.). Further, Nesterov accelerated gradient was proposed to have some prescience about the next update direction (e.g., \cite{havaei_hemis_2016}). To adapt the learning rate to the parameters, Adagrad \cite{duchi2011adaptive} was proposed. However, Adagrad accumulates the squared gradients in the denominator and leads to monotonically decreasing learning rate. RMSprop and Adadelta \cite{zeiler2012adadelta} are proposed to resolve Adagrad's radically diminishing learning rates. Adaptive Moment Estimation (Adam) \cite{kingma_adam:_2014} is an optimizer with momentum and adaptive learning rates. AMSGrad \cite{reddi2019convergence} is a variation of Adam. For MS lesion segmentation, Adadelta (e.g., \cite{birenbaum_longitudinal_2016,brosch_deep_2016,valverde_improving_2017,valverde_one-shot_2018,aslani_deep_2019,mckinley2016nabla}) and Adam (e.g., \cite{kazancli_multiple_2018,roy_multiple_2018,aslani_multi-branch_2018,feng_self-adaptive_2018}) are most widely used. 




\subsection{Prediction and Post-processing}

\paragraph{\bf Prediction.}

For non-FCN methods, segmentation is made by classifying all the candidates extracted from the test image voxel-by-voxel. For FCN methods, 2D networks predict the segmentation slices-by-slice, and 3D methods are able to predict the whole image at once. 
However, Hashemi \emph{et al.} \cite{hashemi_asymmetric_2019} argue border predictions made by FCN are not as accurate as center voxel predictions, so they propose to predict the results patch-by-patch and then fuse predictions using B-spline weighted soft voting, such that border predictions are given lower weights. 
For some methods where data from multiple sources (e.g., views) is used or multiple models are trained, a label aggregation step is necessary. Aslani \emph{et al.} \cite{aslani_deep_2019,aslani_multi-branch_2018} and Zhang \emph{et al.} \cite{zhang_multiple_2019} use majority vote to aggregate labels, but other methods (e.g., STAPLE \cite{warfield2004simultaneous}) can also be considered.

Since the outputs of networks are usually soft predictions (i.e., in the range [0, 1]) indicating the probability of being lesions, the simplest way to get hard predictions is to use a threshold of 0.5 \cite{salehi_tversky_2017}. However, McKinley \emph{et al.} \cite{mckinley_automatic_2020} argue that the scores output by deep networks do not correspond to observed probabilities and are typically overconfident. Brosch \emph{et al.} \cite{brosch_deep_2016} and Roy \emph{et al.} \cite{roy_multiple_2018} attempt to choose the optimal threshold by maximizing the Dice on the training set. 

\paragraph{\bf Post-processing.}

Many methods attempt to remove false positives from the hard segmentation using post-processing strategies. 
A common post-processing approach consists of discarding lesions smaller than a volume threshold (e.g., \cite{brugnara_automated_2020}). Vaidya \emph{et al.} \cite{vaidya2015longitudinal} use pre-built brain templates to remove predicted lesions outside of white matter, but this strategy is problematic for detecting cortical lesions. Kamnitsas \emph{et al.} \cite{kamnitsas_efficient_2017} use an additional stage of machine learning for post-processing, specifically, a fully-connected Conditional Random Field (CRF), a common strategy.
Valverde \emph{et al.} \cite{valverde_improving_2017,valverde_one-shot_2018} use cascaded networks, in which the first network is trained to recall many possible lesions and the second network refines the output of the first network. Specifically, the training data for the second model is balanced between all the lesion voxels and the random selection of misclassified lesion voxels on the first model. Others \cite{kazancli_multiple_2018,xiang_segmentation_2020} use similar strategies.

Nair \emph{et al.} \cite{nair_exploring_2020} present multiple uncertainty estimates based on Monte Carlo (MC) dropout. Then lesions with high uncertainty can be removed. Their results suggest that uncertainty measures allow choosing superior operating points, compared to only using the network’s sigmoid output as a probability.

\subsection{Transfer Learning and Domain Adaptation} 

Transfer learning is an active topic in deep learning, and in the context of MS lesion segmentation, it includes two aspects: 1) using pre-trained models from other domains; 2) applying trained model on different MS datasets (e.g., for clinical use). The latter scenario is also considered as domain adaptation, in which the target task remains the same as the source (i.e., MS lesion segmentation) but the domains (i.e., MRI protocols and therefore image appearance) are different. Multi-task training (Sec. \ref{network}) is also a form of transfer learning.

\paragraph{\bf Pre-training with Other Domains.} The pre-trained blocks (layers) from other domains are typically used to replace the encoder.
Brosch \emph{et al.} \cite{brosch_deep_2016} propose to pre-train the model layer-by-layer with convolutional restricted Boltzmann machines and then apply parameters on both encoder and decoder. Aslani \emph{et al.} \cite{aslani_multi-branch_2018,aslani_deep_2019} use the ResNet50 pre-trained on ImageNet as the encoder. Fenneteau \emph{et al.} \cite{fenneteau_learning_2020} present a self-supervision method to pre-train the encoder to predict the location of an input patch. However, their results illustrate that pre-training is not helpful. Kruger \emph{et al.} \cite{kruger2020fully} pre-train the encoder path with single time point data and then train the entire network with longitudinal data.

\paragraph{\bf Generalization of Trained Models.}
For domain adaptation when a few labeled images are available in the target (new) domain, Ghafoorian \emph{et al.} \cite{ghafoorian_transfer_2017} propose to freeze the first few layers of the model trained on the source domain and fine-tune the last few layers. Their results show that using even just 2 images can achieve a good Dice score. Valverde \emph{et al.} \cite{valverde_one-shot_2018} propose to freeze all convolutional layers and fine-tune the fully connected layers. A single image for re-training could generate segmentation with human-level performance. Weeda \emph{et al.} \cite{weeda_comparing_2019} further test the one-shot learning proposed by \cite{valverde_one-shot_2018} with an independent dataset, and the performance is better than unsupervised methods and is comparable to fully trained supervised methods.

For domain adaptation without any labeled data in the target domain, Baur \emph{et al.} \cite{baur_semi-supervised_2017} propose to add auxiliary manifold embedding loss for utilizing unlabeled data from target domains. The idea is that latent feature vectors that share the same label (for labeled data) or same noisy prior (unlabeled data) should be similar, and otherwise differ from each other. Baur \emph{et al.} \cite{baur_fusing_2018} propose to train an auto-encoder for unsupervised anomaly detection in the target domain, and use this unsupervised model to generate artificial labels for jointly training a supervised model with labeled data from the source domain.
Ackaouy \emph{et al.}  \cite{ackaouy_unsupervised_2020} propose a method to perform unsupervised domain adaptation with optimal transport. In the deep learning context, their strategy is implemented as two losses, which ensures the heavily connected source samples and target samples to have similar representations in the latent space and the output, while maintaining good segmentation performance. 

Billast \emph{et al.}  \cite{billast_improved_2020} present domain adaptation with adversarial training. The discriminator is trained to discriminate whether the two input segmentations are from the same scanner, so that the generator learns to map scans from different scanners to the same latent space and thus produce a consistent lesion segmentation. 
Aslani \emph{et al.}  \cite{aslani_scanner_2020} propose a similar idea with a regularization network predicting the feature domain. They use a combination loss of Pearson correlation, randomized cross-entropy and discrete uniform to encourage the latent features to be domain agnostic.
Varsavsky \emph{et al.} \cite{varsavsky2020test} combine domain adversarial learning and consistency regularization, which enforces invariance to data augmentation.


\subsection{Methods for Subtypes of MS Lesions}
Most of the methods we have discussed are proposed to segment white matter lesions. Among these, some pipelines focus on detecting contrast-enhancing (CE) lesions since these are indicative of active disease. Gadolinium (Gad) is commonly used in the context of MS.
Durso-finley \emph{et al.} \cite{durso-finley_saliency_2020} and Coronado \emph{et al.} \cite{coronado_deep_2020} present to detect Gad lesions using pre- and post-contrast T1-weighted images.
Brugnara \emph{et al.} \cite{brugnara_automated_2020} propose a network to detect both CE lesions and T2/FLAIR-hyperintense lesions and report the performance separately. 
On the other hand, radiological monitoring of disease progression also requires detecting new and enlarging T2w lesions, which can be explored by longitudinal approaches (Sec.~\ref{sec:multi-sources}, multi-timepoints).
La Rosa \emph{et al.} \cite{la_rosa_shallow_2018} propose to detect the early stage lesions by combining deep neural networks with a shallow model (supervised k-NN with partial volume modeling). 



Cortical lesions are also important in MS \cite{calabrese2009cortical}. La Rosa \emph{et al.} \cite{la_rosa_multiple_2020} use simplified U-Net to detect both cortical and white matter lesions at 3T MRI.
To achieve this, they utilize 3T 3D-FLAIR and magnetization-prepared 2 rapid acquisition with gradient echo (MP2RAGE).
They further use 7T MRI (7T MP2RAGE, T2*w echo planar imaging, T2*w gradient recalled echo) for cortical lesion segmentation, which has higher resolution and SNR than 3T \cite{la_rosa_automated_2020}.

\section{Comparison of Experiments and Results}

\subsection{Datasets}

\paragraph{\bf Public Datasets.}

Currently, the challenge datasets, including the MICCAI 2008\footnote{\url{http://www.ia.unc.edu/MSseg}} \cite{styner_2008a_2008}, ISBI 2015\footnote{\url{https://smart-stats-tools.org/lesion-challenge}} \cite{carass_longitudinal_2017}, MICCAI (MSSEG) 2016\footnote{\url{https://portal.fli-iam.irisa.fr/msseg-challenge/overview}} \cite{commowick_objective_2018} challenges, are widely used. The dataset descriptions can be found on the respective websites. The first two challenges are still (as of November 2020) accepting segmentation submissions on their test dataset, which provides objective comparisons between state-of-the-art MS segmentation methods. 
Lesjak \emph{et al.} \cite{lesjak_novel_2018} provided a novel public dataset\footnote{\url{http://lit.fe.uni-lj.si/tools}}, for which three expert raters performed segmentation of WM lesions and reached consensus by several joint sessions. 
They illustrated that the consensus-based segmentation have better consistency than a single rater's segmentation.
It is worth noting that all these public datasets only delineate white matter lesions.

\paragraph{\bf Private Datasets.} 
Using private datasets makes it difficult to compare algorithms but has the advantage of including more subjects than currently available in public datasets.
Some proprietary datasets can be of a quite large scale (e.g., 6830 multi-channel MRI \cite{durso-finley_saliency_2020}). For such large datasets, the ``ground truth'' labels are usually created by automated or semi-automated algorithms and corrected by experts \cite{la_rosa_shallow_2018,durso-finley_saliency_2020}. 

Narayana \emph{et al.} \cite{narayana_deep-learning-based_2020} considered the effect of training data size for training the neural networks. They argue that at least 50 image volumes are necessary to get a meaningful segmentation. But this work does not mention the data augmentations and other advanced techniques for training a network. Based on the results of the ISBI 2015 challenge \cite{carass_longitudinal_2017}, human-level performance can be achieved by state-of-the-art algorithms with only about 20 images for training. 


\subsection{Evaluation Metrics}

In the task of MS lesion segmentation, the commonly reported metrics include:
Dice similarity coefficient (DSC), Jaccard coefficient, absolute volume difference (AVD), average symmetric surface distance (ASSD/SD), true positives rate (TPR, sensitivity, recall), false positives rate (FPR), positive predictive value (PPV, Precision), lesion-wise true positives rate (LTPR) and false positives rate (LFPR).

The above metrics are individually calculated based on each image. Then, the results are aggregated and reported. In addition to reporting mean and standard deviation values, the Wilcoxon signed-rank test is used to statistically test performance differences between methods. Precision-Recall (PR) curve is suitable for evaluating the performance of the highly unbalanced dataset. Receiver Operating Characteristic (ROC) curve is also used (e.g., \cite{nair_exploring_2020}). The area under curve (AUC) for these curves is a common aggregation metric. 
Further, considering the relationship between the model performance and lesion volume, some works divide lesions into groups of different sizes and calculated the metrics (e.g., \cite{wang_ensemble_2018}). Volumes of lesions estimated and the ground truth segmentation can be shown in the correlation (e.g., \cite{valverde_improving_2017,kazancli_multiple_2018,roy_multiple_2018,aslani_multi-branch_2018,duong_convolutional_2019}) and Bland-Altman (e.g., \cite{la_rosa_shallow_2018,mckinley_simultaneous_2019}) plots.
A more systematic analysis of algorithm performance can differentiate between correctly detected lesions, nearby lesions merged into one or a single lesion split into many, as well as characterize the performance as a function of lesion size \cite{oguz2017dice,carass2020evaluating}.

\subsection{Results}

As previously mentioned, MICCAI 2008 \cite{styner_2008a_2008} and ISBI 2015 \cite{carass_longitudinal_2017} are still accepting submissions and providing the evaluation results on the test dataset, thus serving as objective benchmarks. In this survey, we compare the state-of-the-art methods that have evaluated their performance on these datasets in Table \ref{tab:miccai2008} and Table \ref{tab:isbi}.

\begin{table}[htbp]
\centering
\caption{Results on the ISBI 2015 challenge test set. All metrics in percent. DSC: Dice; PPV: Precision; TPR: true positives rate; LTPR: lesion-wise TPR; LFPR: lesion-wise false positives rate; VD: volume difference; SC: total weighted score of other metrics. Code: links to code repositories,if available.}
\label{tab:isbi}
\begin{tabular}{l @{\hspace{0.5cm}}  l @{\hspace{0.38cm}} c @{\hspace{0.38cm}} c @{\hspace{0.38cm}} c @{\hspace{0.38cm}} c @{\hspace{0.38cm}} c @{\hspace{0.38cm}} c @{\hspace{0.38cm}} c @{\hspace{0.2cm}} }
\toprule
 & SC & DSC & PPV & TPR & LFPR & LTPR & VD & Code\\
\midrule
Zhang \emph{et al.} \cite{zhang_multiple_2019} & 93.21 & 64.3 & 90.8 & 53.3 & 12.4 & 52.0 & 42.8 &
\href{https://github.com/MedICL-VU/LesionSeg}{Yes}
\\
Isensee \emph{et al.}  \cite{isensee_nnu-net_2019} & 92.87 & 67.9 & 84.7 & 60.5 & 15.9 & 52.2 & 36.8 &
\href{https://github.com/MIC-DKFZ/nnUNet}{Yes}
\\
Hu \emph{et al.} \cite{hu_acu-net_2020} & 92.61 & 63.4 & 86.9 & 52.6 & 13.4 & 48.2 & 39.7 & No \\
Hashemi \emph{et al.} \cite{hashemi_asymmetric_2019} & 92.49 & 58.4 & 92.1 & 45.6 & 8.7 & 41.3 & 49.7 & No \\
Feng \emph{et al.} \cite{feng_self-adaptive_2018} & 92.41 & 68.2 & 78.2 & 64.5 & 27.0 & 60.0 & 32.6 & No \\
Denner \emph{et al.} \cite{denner_spatio-temporal_2020} & 92.12 & 64.3 & 85.9 & 54.5 & 19.5 & 47.1 & 38.6 &
\href{https://github.com/StefanDenn3r/Spatio-temporal-MS-Lesion-Segmentation}{Yes}
\\
Aslani \emph{et al.} \cite{aslani_multi-branch_2018}& 92.12 & 61.1 & 89.9 & 49.0 & 13.9 & 41.0 & 45.4 & No \\
Valverde \emph{et al.} \cite{valverde_improving_2017}& 91.33 & 63.0 & 78.7 & 55.5 & 15.3 & 36.7 & 33.8 &
\href{https://github.com/sergivalverde/cnn-ms-lesion-segmentation}{Yes}
\\
Roy \emph{et al.} \cite{roy_multiple_2018} & 90.48 & 52.4 & 86.6 & N/A & 11.0 & N/A & 52.1 &
\href{https://www.nitrc.org/projects/flexconn}{Yes}
\\
Valverde \emph{et al.} \cite{valverde_one-shot_2018}\footnotemark & 90.32 & 57.7 & 83.1 & 47.5 & 18.9 & 29.7 & 44.6 &
\href{https://github.com/sergivalverde/nicMSlesions}{Yes}
\\
Birenbaum \emph{et al.} \cite{birenbaum_longitudinal_2016} & 90.07 & 62.7 & 78.9 & 55.5 & 49.8 & 56.8 & 35.2 & No \\
\bottomrule
\end{tabular}
\end{table}
\footnotetext{trained on other datasets and fine-tune with one sample from this dataset.}

\begin{table}[htbp]
\centering
\caption{Results on the MICCAI 2008 challenge test set. Subscript 1: UNC Rater; 2: CHB Rater. All metrics in percent, except SD in millimeters. VD: volume difference; SD: surface distance; TPR: true positives rate; FPR: false positives rate; SC: total weighted score of other metrics.}
\label{tab:miccai2008}
\begin{tabular}{l @{\hspace{0.5cm}}  l @{\hspace{0.30cm}} c @{\hspace{0.30cm}} c @{\hspace{0.30cm}} c @{\hspace{0.30cm}} c @{\hspace{0.30cm}} c @{\hspace{0.30cm}} c @{\hspace{0.30cm}} c @{\hspace{0.30cm}} c @{\hspace{0.30cm}}}
\toprule
 & SC & VD$_1$ & SD$_1$ & TPR$_1$ & FPR$_1$ & VD$_2$ & SD$_2$ & TPR$_2$ & FPR$_2$\\
\midrule
Valverde \emph{et al.} \cite{valverde_improving_2017} & 87.1 & 62.5 & 5.8 & 55.5 & 46.8 & 40.8 & 5.2 & 68.7 & 46.0 \\
Brosch \emph{et al.} \cite{brosch_deep_2016} & 84.0 & 63.5 & 7.4 & 47.1 & 52.7 & 52.0 & 6.4 & 56.0 & 49.8\\
Havaei \emph{et al.} \cite{havaei_hemis_2016} & 83.2 & 127 & 7.5 & 66.1 & 55.3 & 68.2 & 6.6 & 52.3 & 61.3\\
\bottomrule
\end{tabular}
\end{table}

From the results, we observe that 3D and 2.5D methods seem to outperform 2D approaches with the development of GPUs. As in Table \ref{tab:isbi}, U-Net-based methods (\cite{zhang_multiple_2019,isensee_nnu-net_2019,hu_acu-net_2020,hashemi_asymmetric_2019,denner_spatio-temporal_2020,aslani_multi-branch_2018}) tend to perform better than non-FCN CNN-based (\cite{birenbaum_longitudinal_2016,valverde_improving_2017}) and non-U-Net FCN-based (\cite{roy_multiple_2018}) methods.

\section{Conclusion}

In this survey, we explored the advances in different components of supervised CNN MS lesion segmentation methods. Among these, topics including attention mechanism, network designs to combine information from multiple sources, loss functions to handle class imbalance, and domain adaptation are of interest for many researchers. 

\paragraph{Acknowledgements.} This work was supported, in part, by the NIH grant R01-NS094456 and National Multiple Sclerosis Society award PP-1905-34001.

\bibliography{ref} 
\bibliographystyle{splncs04}

\end{document}